# The DARPA TWITTER BOT CHALLENGE[1]

V.S. Subrahmanian, University of Maryland, vs@cs.umd.edu

Amos Azaria, Carnegie Mellon University and Sentimetrix, amos.azaria@gmail.com

Skylar Durst and Vadim Kagan, Sentimetrix, skylar@sentimetrix.com, Kagan@sentimetrix.com

Aram Galstyan, Kristina Lerman, and Linhong Zhu, Univ. of Southern California galstyan@isi.edu, lerman@isi.edu, linhong@isi.edu

Emilio Ferrara, Alessandro Flammini, and Filippo Menczer, Indiana University Emilio.ferrara@gmail.com, aflammin@indiana.edu, fil@indiana.edu

*Additional Authors*: Andrew Stevens (Sentimetrix), Alexander Dekhtyar (CalPoly and Sentimetrix), Shuyang Gao (USC), Tad Hogg (Institute for Molecular Manufacturing) Farshad Kooti (USC), Yan Liu (USC), Onur Varol (Indiana University), Prashant Shiralkar (Indiana University), Vinod Vydiswaran (University of Michigan), Qiaozhu Mei (University of Michigan), Tim Hwang (Pacific Social).

## ABSTRACT

A number of organizations ranging from terrorist groups such as ISIS to politicians and nation states reportedly conduct explicit campaigns to influence opinion on social media, posing a risk to democratic processes. There is thus a growing need to identify and eliminate "influence bots" - realistic, automated identities that illicitly shape discussion on sites like Twitter and Facebook - before they get too influential. Spurred by such events, DARPA held a 4-week competition in February/March 2015 in which multiple teams supported by the DARPA Social Media in Strategic Communications program competed to identify a set of previously identified "influence bots" serving as ground truth on a specific topic within Twitter. Past work regarding influence bots often has difficulty supporting claims about accuracy, since there is limited ground truth (though some exceptions do exist [3,7]). However, with the exception of [3], no past work has looked specifically at identifying influence bots on a specific topic. This paper describes the DARPA Challenge and describes the methods used by the three top-ranked teams.

**Keywords: Social Networks, Intrusion/Anomaly Detection and Malware Mitigation**

---

I. **MOTIVATION**

According to a recent Twitter SEC filing, approximately 8.5% of all Twitter users are bots. While many of these bots have a commercial purpose such as spreading spam, some are "influence bots" – bots whose purpose is to shape opinion on a topic. This poses a clear danger to freedom of expression. For instance, the terrorist group ISIS used online social media to spread radicalism [24] by influencing youth to embrace their cause. On the political front, [17] asserts that Russia waged a social media disinformation campaign in the aftermath of Russian actions in the Ukraine. In [27], a class in Denmark showed that they could build social bots that had surprisingly large influence.

Twitter bots [1] include:

- *Spambots* spread spam on various topics.
- *Paybots* illicitly make money. Some paybots copy tweet content from respected sources like @CNN but paste in micro-URLs that direct users to sites that pay the bot creator for directing traffic to the site.
- *Influence* bots try to influence Twitter conversations on a specific topic. For instance, some politicians have been accused of buying influence on social media.

In the first quarter of 2015, DARPA conducted the "Twitter Bot Detection Challenge": a 4-week competition to test the effectiveness of influence bot detection methods developed under the DARPA Social Media in Strategic Communications (SMISC) program. The challenge was to identify influence bots supporting a pro-vaccination discussion on Twitter. There is a vocal "anti-vaccination" community both on the Internet and on Twitter [2] and other relevant social media.

Since the challenge focused on identifying influence bots seeking to diffuse a sentiment $s$ on a topic $t$, competitors had to:

1. Separate influence bots from other types of bots.
2. Separate influence bots about topic $t$ from those about other topics.
3. Separate influence bots about topic $t$ that sought to spread sentiment $s$ from those that were either neutral or that spread an opposite sentiment.

6 teams (University of Southern California, Indiana University, Georgia Tech, Sentimetrix, IBM, and Boston Fusion), competed to discover 39 pro-vaccination influence bots. The teams did not know the number of bots. Table 1 below summarizes the final results of the competition that was won by Sentimetrix which beat other teams by 6 days with 39 of 40 guesses being correct. USC achieved the best accuracy (39 of 39 correct guesses) of all teams.

|  | Misses | Hits | Guesses | Accuracy | Speed | Final Score |
|---|---|---|---|---|---|---|
| Sentimetrix | 1 | 39 | 40 | 38.75 | 12 | 50.75 |
| USC | 0 | 39 | 39 | 39 | 6 | 45 |
| DESPIC | 7 | 39 | 46 | 37.25 | 6 | 43.25 |
| IBM | 4 | 39 | 43 | 38 | 5 | 43 |
| B. Fusion | 9 | 39 | 48 | 36.75 | 5 | 41.75 |
| G. Tech | 56 | 38 | 94 | 24 | 0 | 24 |

**Table 1. Results of the DARPA Twitter Bot Challenge.** The "Accuracy" Column is the value (*h*-0.25*m*) where *h* is the number of hits (correct guesses) and *m* is the number of misses (incorrect guesses). The "Speed" column equals the number of days remaining in the Challenge when the team had discovered all bots. (DESPIC is the Indiana University team which also included University of Michigan). For each team $t$, $FinalScore(t) = Hits(t) - 0.25 * Misses(t) + Speed$.

This paper describes how the 3 top teams (Sentimetrix, USC, and University of Indiana) achieved their results.

II.        SETUP OF THE DARPA SMISC TWITTER BOT CHALLENGE

In Fall 2014, Pacific Social Inc. logged records of an influence competition independently taking place on Twitter on the use of influence bots in combating misinformation online, specifically around anti-vaccine activists on Twitter. Using ground truth on the teams and the bot accounts they operated, they developed a synthetic Twitter environment with a simulated Twitter API that played back a partially redacted set of their data. This data consisted of:

- User Accounts: 7,038 accounts.
- User Profiles: Redacted user profiles with Twitter-like format: user image, URL, number of friends and followers, plus a short user bio.
- Tweets: A time-stamped tweet data set for each user (4,095,083 tweets in all).
- Network Data: Weekly network snapshots consisting of *(from_user, to_user, timestamp, weight)* tuples. A tuple's weight was "0" if "from_user" unfollowed "to_user", and was "1" otherwise. There were 17,503 users for whom partial network data was provided – the data included user-ids that were not present in the user accounts.

Once the challenge started, teams could submit guesses to a webserver which would immediately provide right/wrong information. Team scores were computed as follows:

- A team received one point for each correct guess.
- A team lost ¼ point for each false positive (called "miss" in the above table).
- A team that guessed *all* the bots *d* days before the Challenge ended received *d* extra points.

The third scoring clause provided a bonus for speedy guessing. Identifying real world bots early is important in order to counteract an influence campaign. This is particularly important in adversarial situations (e.g. the ISIS's social media campaign [24] or attempts to inappropriately influence an election).

Sentimetrix guessed all bots on Day 16 of 28, receiving 12 bonus points. Of 40 guesses made, only one was wrong, for a final score of 50.75. The speed bonus gave Sentimetrix an almost one week edge over the nearest competitors (USC and Indiana University). Both USC and Indiana found all bots 6 days later. USC had perfect precision, while Indiana had 7 erroneous guesses.

### III. BOT DETECTION APPROACHES

*All three winning parties found that machine learning techniques alone were insufficient because of lack of training data. However, a semi-automated process that included machine learning proved useful.*

No team started with unsupervised learning. USC used unsupervised outlier detection in conjunction with other evidence. Indiana and Sentimetrix used clustering algorithms later, while Indiana also tried an online prediction strategy. No team found existing Sybil detection methods [5,6] useful. All teams benefited from previous influence-bot studies [1,3,9,16].

### III.A Creating a Training Set

All but one team used past work to build a profile *Prof(u)* of user *u* [1,3]. Sentimetrix's SentiBot [1] identified influence bots from the 2014 Indian election [4] over 10 months, amassing a dataset exceeding 17M users, 25M tweets, and 45M edges. [3]'s dataset (collected from 60 social honeypots deployed over 7 months involving 42K users) was used by two teams to learn models separating bots from non-bots. Irrespective of the method used, the features below were of interest.

***Tweet Syntax.*** This included the following features:

- Does the user post tweets whose syntax is similar to the natural language generation program Eliza [7] and auto-generation of language [8]?
- Average number of hashtags, user mentions, links, and special characters in tweets.
- Average number of retweets by the user.
- Are the tweets geo-enabled?
- Percentage of tweets ending with punctuation, hashtag, or link – the intuition is that such tweets might be automatically generated.

***Tweet Semantics.*** This category included the following features:

- Number of user posts related to vaccination.
- User's average sentiment score (on the topic "vaccination") in vaccination-related tweets.
- Measures of contradiction in the user's posts on vaccination-related tweets using functions such as Contradiction Rank [1] which measures variation between the sentiment of the user across a set of topics and the sentiments of his neighbors on the same topics.

- Positive (resp. Negative) Sentiment Strength [1] measuring the average sentiment strength of the user's positive (resp. negative) tweets.
- Most frequent topics that the user tweets about.
- Number of languages in which tweets were generated. The rationale was that accounts posting tweets in many languages may be bots.
- Sentiment inconsistency. [1] noticed that paybots often copy a link from a popular Twitter user and then replace the micro-URLs from the original Twitter post with a spurious link to a site where the paybot owners were paid for generating views. This feature analyzes whether sentiment in the tweet content varied significantly (w.r.t. the topic of vaccination) as compared to sentiment in a URL embedded within the tweet.

Teams used NLP tools such as Latent Dirichlet Allocation [10] for topic detection, and AVA [11] and OASYS [12] to assign vaccination sentiment scores in the [-1,+1] range. USC used a tri-partite graph clustering approach to infer tweet-level and user-level sentiment [19].

***Temporal Behavior Features.*** This category tracked how user sentiments changed over time.

- Did the user's sentiment flip-flop over time [1]?
- Variance in tweet sentiment over time [1]. This enabled Sentimetrix to identify users who had an explicit infiltration strategy of posting anti-vaccination tweets to engage the anti-vaccination community and later switching to a pro-vaccination stance.
- Entropy of inter-tweet time distribution [9,20]. The rationale was that algorithmic tweeting should have some temporal regularities that are reflected in relatively low entropy of the corresponding distribution.
- Predictability of tweet timing based on a transfer entropy approach [21].
- The duration of the longest session by a user without any short (5-minute or 10-minute) breaks – clearly, users that have a session lasting a day without any breaks are not likely to be humans.
- Average number of tweets per day – as in the previous case, if this number is large, it increases the probability that the user is a bot.
- Percentage of dropped followers. What was the percentage of "unfollows" compared to the percentage of "follows"? For instance, a user who dropped a lot of followers compared to the number of people he was following may be anomalous.
- Signal-to-noise ratio (SNR). Indiana tracked the ratio of mean to standard deviation, min, max and entropy of these values to detect abrupt changes in users' metadata (followers, followees, posts, etc.).

***User Profile Features.*** User profile features considered aspects such as:

- Did the user's profile have a photo? If so, was it from a stock image database?
- Did the user's profile have an associated URL? If so, did the URL have a clone elsewhere? A URL that was a clone of some other URL increased the level of suspicion of the user.

- Did the user's Twitter name look auto-generated? Sentimetrix generated several heuristics for such tests, e.g. by comparing screen-names with user-names after splitting on spaces/underscores and looking for common substrings.
- Number of posts/retweets/replies/mentions.
- Number of followers/followings.
- Number of sources used by the user such as: mobile applications, desktop browsers, or "null" for missing sources.
- GPS coordinate availability for user's tweets.
- Similarity of user profile to known bots (measured using Jaccard similarity by Sentimetrix and Cosine Similarity by Indiana).

*Network Features.* Teams used network-related sources such as:

- Average Deviation of user sentiment scores from those of his followers and followings.
- In and Out degree centrality.
- Average clustering coefficient of retweet and mention network associated with each user.
- Pagerank and between-ness centrality of users in both retweet and mention networks.
- Variables related to star and clique networks associated with users.
- Number of known bots followed by a user – a user following several known bots is more likely to be a bot.
- Number/Percentage of bots in the cluster that a user belonged to –if a clustering algorithm places the user in a cluster with many bots, he is more likely to be a bot.

Different teams considered different numbers of features. Some teams added features once the Challenge started and some bots had been discovered. Sentimetrix started with 66 features (which increased to 175 by the end), Indiana used 98, and USC used 47.

Teams were then able to use insights from past work [1,3] to identify a small number of initial bots by manually inspecting suspicious accounts. For instance, Sentimetrix identified 4 initial bots this way and then used clustering and network analysis (see Section III.B) to identify 25 more bots. They then used Support Vector Machines (SVM) to predict the remaining 10 bots using the features described in this section and the analytic tools of Section III.B. Similarly, USC detected the first 4 bots by combining outlier detection with content analysis and manual inspection. The next 21 bots were detected using a combination of network analysis (e.g., connection to known bots), content and sentiment analysis, and semi-supervised clustering of users based on the above-described features.

**III.B FEATURE ANALYSIS**

The feature data for each user in Section III.A was periodically updated. Sentimetrix automatically updated its feature data overnight. All three teams writing this paper had internal dashboards that allowed team members to navigate and display competition data. Teams used multiple analytical tools for competition bot prediction. We now describe these components in greater detail.

**III.B.i Bot Analysis Dashboards**

All three teams used bot analysis dashboards. Sentimetrix's Bot Analysis Dashboard showed details on every single user account in the DARPA Twitter Bot Challenge data. Figure 1 below shows the main screen, providing the analyst with a birds-eye view of all users.

Each user has a flag telling analysts if he is "active" and a bot/human/other label. Some users are marked as "bots" (these are users identified as bots by the system and confirmed by human inspection). Other users may be marked "human" and/or with some flags (e.g. "profile image mismatch") suggesting that something suspicious is going on. These labels were used during clustering and SVM training, allowing us to discover additional groups of users.

A user summary described how complete the user profile is and a description of the person (e.g. follower-followee ratio). A number of additional variables are associated with each user – but these are not visible in Figure 1 (as those columns extend beyond the right edge of Figure 1). Many of these variables cue an analyst about whether a person is suspicious. The Sentimetrix Dashboard allows analysts to query the profiles, sort them in descending order of columns (see top of Figure 1).

![Sentimetrix Dashboard screenshot showing the All Users view with columns for Image, View, Active, Label, Tweeter Id, Screen Name, Name, Description, Profile Complete %, and Follower. Rows include CarrieWoolf, susan_east, rhondapgranger, MildredMason19, NicoleSGeorge, Good_Aftenoon, RStationery, senpai156, JanetteSohm, real_sex_story, DefendingBeef, ZPerrington, onetello7.]

**Figure 1. Sentimetrix Dashboard to view Twitter user information.**

The Sentimetrix Dashboard also provides information specific for each user (shown when a particular user is clicked). This information includes snapshots of the network at different times, which were updated as the competition proceeded.

Figure 2 shows the details of a Twitter user (*gunslinger_mk1*) which had no profile image for a while – and then at some point during the competition, his profile image was updated. Details of the background image used are also shown. The top left of the screen shows that the system classified *gunslinger_mk1* as a bot.

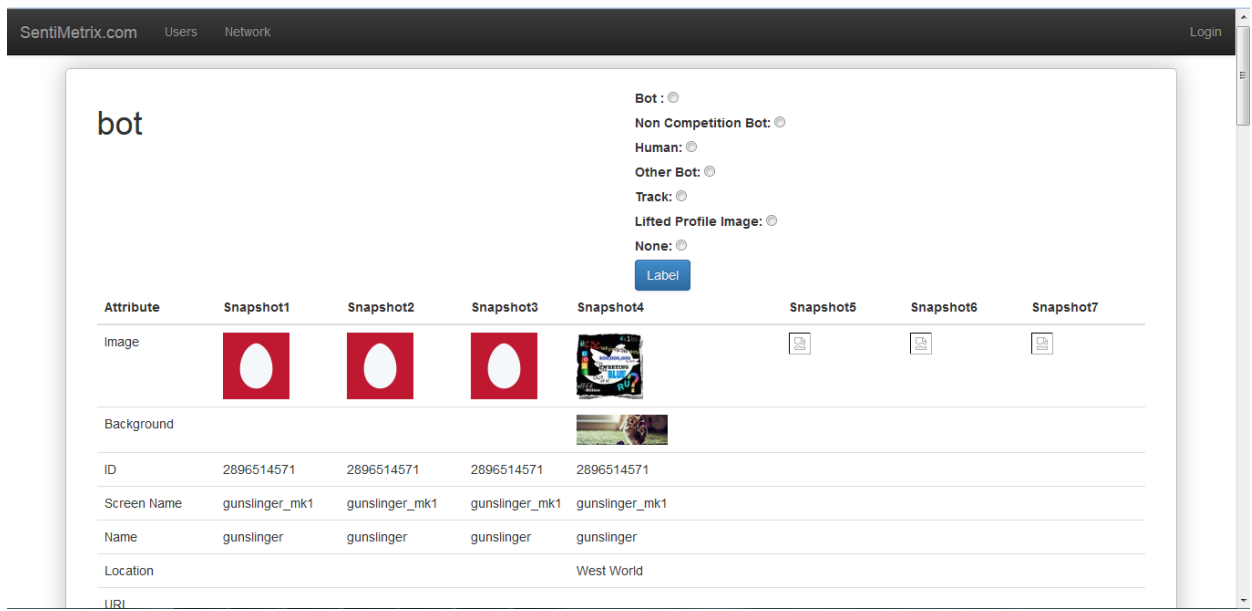

**Figure 2. Top of detailed screen about a Competition Bot**

Sentimetrix used other screens (not shown) to further help identify bots. Using these capabilities, the analyst can quickly identify suspect users and flag them as bots.

**III.B.ii Bot Analysis Algorithms**

The three teams used several bot analysis algorithms, some of which are described in this section.

*Hashtag Co-Occurrence Network.* Starting from the provided list of vaccine-related hashtags, Indiana constructed a hashtag co-occurrence network – nodes represent unique hashtags and edges between two nodes are weighted by the number of times these two hashtags co-occur in a tweet (Fig.3). Indiana used these hashtag co-occurrence networks to identify other campaign-related hashtags to enrich the list of competition-relevant keywords. These were later used to separate users into pro- and anti-vaccine categories. The proportion of tweets users posted containing any of these hashtags resulted in a strongly predictive feature.

**Figure 3. Hashtag co-occurrence network**

*Distance Measures.* The Indiana team identified additional bots by computing the cosine similarity between users and known bots. Figure 4 shows the kernel density estimation of the pairwise cosine distance between pairs of feature vectors characterizing two bots, compared to bot-human pairs. The distances between bot pairs are much smaller than bot-human pairs. The bot-bot distance exhibits a bimodal distribution that reflects the presence of two types of bots designed by two teams. Sentimetrix achieved similar success using Jaccard distance.

**Figure 4. Kernel density estimation of the cosine Distance between bot-bot pairs (blue) and bot-human pairs.**

*Online Prediction*. Indiana also adopted a multi-arm bandit based online prediction strategy. "Arms" were initialized with a set of binary classifiers, with a hedge-like algorithm [18] that decided what arm to pull next. Each user account got a prediction score between 0 and 1 assigned by each arm (the higher the score, the higher the likelihood of being a bot). The hedge-like algorithm initially assigned uniform weights to all arms, and then used a multiplicative scheme to update the arm weights. After each round, it produced a final "bot score" for each user as a weighted average of the prediction scores of each arm. It then selected the account with the highest bot score as the next guess. Upon receiving the feedback score $x$ (positive or negative), the weight of all classifiers were multiplied by a factor of $e^{x \times f_j}$ where $f_j$ is classifier $j$'s prediction score for that guessed account. Thus, accurate classifiers ("arms") gradually gained weight, while inaccurate ones lost weight.

*Outlier Detection.* USC and Sentimetrix assumed that it would be inefficient for bot designers to handcraft bots one at a time – they both expected that one program would generate a number of bots by varying one or more parameters in the bot creation algorithm. Of course, a number of such "bot creation" programs could be deployed by real world bot developers. Because this would lead to similarities amongst bots created by the same program, two detection methods were used for this.

USC first applied orthogonal Non-Negative Matrix Factorization (NMF) [14] to the features in Section III.A to find a low-dimensional vector representation of each user. They then used a clustering-based outlier detection algorithm, to find outliers in this low-dimension latent space. USC performed micro-level clustering via 2 approaches. (i) They used the same feature representation for outliers, and then re-applied NMF to cluster outliers. (ii) They created a similarity graph of outliers using KNN search and then used modularity maximization based community detection [22] to cluster similar outliers. USC's analysis indicates that all the confirmed bots were reported as outliers, yielding a recall score of 1.0. USC also did not make a single false positive, resulting in the only perfect accuracy score in the competition. In contrast to USC, Sentimetrix used the well-known DBSCAN [15] algorithm to generate clusters and then prioritized which users in these clusters were likely to be bots through an analysis of their features and the similarities between those features and features of other known bots.

## IV. OVERALL FRAMEWORK FOR DETECTING INFLUENCE BOTS

*Because new bots will be generated by adversaries who use different (and increasingly sophisticated) bot generation methods, we believe that machine learning by itself would be inadequate.* Our experience is that bot detection is a semi-automated process that builds on four broad techniques: inconsistency detection and behavioral modeling, text analysis, network analysis, and machine learning. Bot detection requires the carefully designed workflow given below, with strong supporting software.

- **Step 1 (Initial Bot Detection).** The first step is to identify a few bots. Team members used four broad classes of cues to uncover a first set of initial bots –*heuristics* (e.g. Indiana University looked for bots that used stock images for profile photos), *behaviors* (e.g. USC and Sentimetrix looked at number of tweets posted over extended periods of times), *linguistic cues* (e.g. Sentimetrix looked for Eliza-style tweets [7] and tweets that terminated with unusual grammar), and *inconsistencies* (e.g. photo showing a bearded older man, while the Twitter handle is MaryJones17 – a user who mostly talks about college).
- **Step 2 (Clustering, Outliers, and Network Analysis).** Though only a few simple bots may be found in Step 1, they are very valuable. Bots connect to each other to inflate follower counts and to increase retweets. As most bot developers write pieces of code that vary parameters in order to generate bots, the shared parameters may create clusters – so clusters containing known bots may include other bots. Sentimetrix exploited these properties to find numerous bots, while USC used outlier analysis to find bots that are distant from all clusters. USC used local ego-networks of known bots to get some insights about the structural connectivity patterns of bots, in order to generate more candidates for guesses.
- **Step 3 (Classification/Outlier Analysis).** Once a certain number of bots and humans are found, we can identify other bots using standard classifiers. For instance, once 29 bots had been found, Sentimetrix used support vector machines to immediately find the remaining 10 bots.

A major problem faced by the top two teams (Sentimetrix and USC) was in determining when to stop guessing. Both teams stopped guessing when they were unable to find any more credible bots to guess. As DARPA announced that two teams had found all the bots immediately after the fact, other teams could guess the number of bots by examining the Sentimetrix and USC scores though we do not know if this was in fact done.

*Underlying these themes is the fact that the system needs to be semi-supervised.* All teams used human judgement to augment automated bot identification processes. Interfaces that easily explain why a particular Twitter account is considered a bot are particularly important. These interfaces must include effective visualizations that highlight the top suspect accounts and explain why they are suspicious. Such interfaces must allow analysts to provide feedback – and take that feedback into account to improve detection accuracy.

## V. CONCLUSION AND FUTURE WORK

Bot developers are becoming increasingly sophisticated. Over the next few years, we can expect a proliferation of social media influence bots as advertisers, criminals, politicians, nation states, terrorists, and others try to influence populations.

As influence bots get more sophisticated, we need to significantly enhance the analytic tools that help analysts detect influence bots. The first step is the full automation of the process described in Section IV. The "Initial Bot Detection" Step needs to be supported by a large collection of tools that systematically cover the search space. The "Clustering, Outliers, and Network Analysis" stage will need to present a toolbox that detects additional bots by looking at clusters of users and outliers. Suspects detected by different algorithms will need to be merged into a single "suspect" list. Powerful visualization methods are needed throughout – to show suspect bots to analysts and to explain why they are suspicious. Once sufficiently many bots have been discovered, along with benign accounts, traditional classifiers can generate additional candidate bots.

While the methods described here were developed for detecting fully automated bots, we believe that those methods can be used for detecting human-orchestrated influence operations as well. Such ongoing influence campaigns include the terrorist group ISIS's attempts to recruit terrorists [24,28] and the reported use of social media by Russia in the context of the ongoing conflict in the Ukraine [17].

***Acknowledgement.*** We thank DARPA and the US Army for funding this project under contracts W911NF-12-C-0026 and W911NF-12-1-0034.

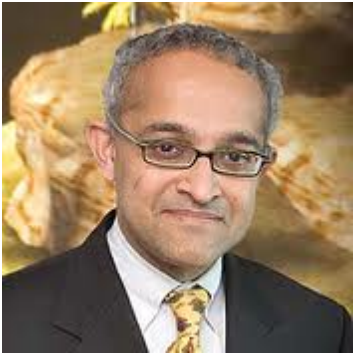

V.S. Subrahmanian is a Professor of Computer Science, a past Director of the University of Maryland Institute for Advanced Computer Studies, and a founder of Sentimetrix.

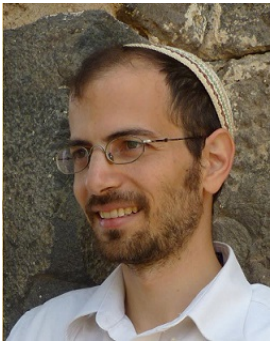

Amos Azaria is a postdoctoral researcher at Carnegie Mellon University in the Machine Learning department.

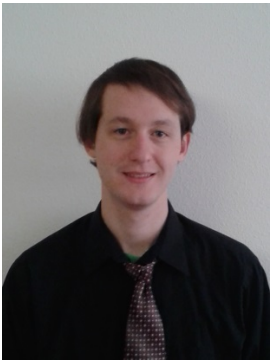

Skylar Durst is a graduate student of Computer Science at the California Polytechnic State University (SLO) and full-time data architect at Sentimetrix

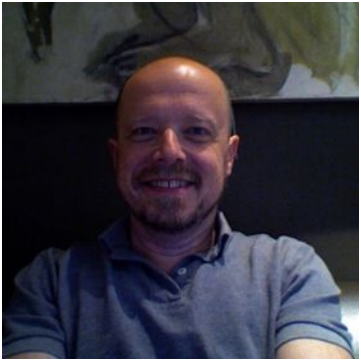

Vadim Kagan is a technologist with over 30 years of experience in building large-scale systems; he is a founder and president of SentiMetrix

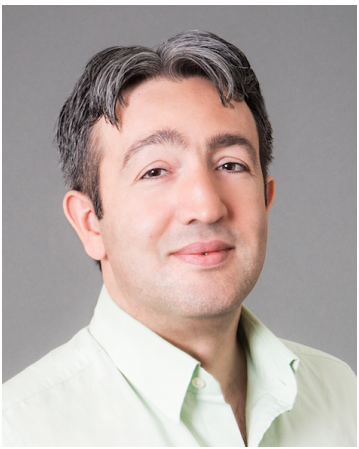

Aram Galstyan is a Project Leader at the USC Information Sciences Institute and a Research Associate Professor of Computer Science at USC

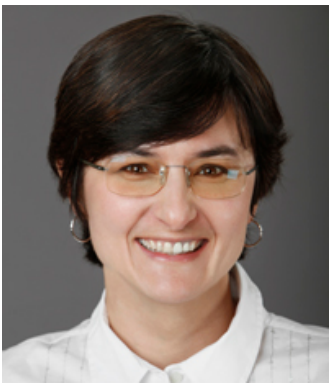

Kristina Lerman is a Project Leader at the [Information Sciences Institute](#) and holds a joint appointment as a Research Associate Professor in the [USC](#) Computer Science Department.

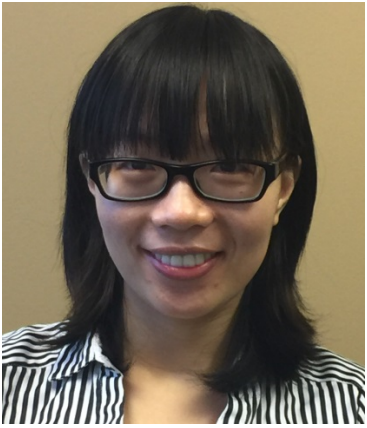

Linhong Zhu is a Computer Scientist at the Information Sciences Institute, University of Southern California.

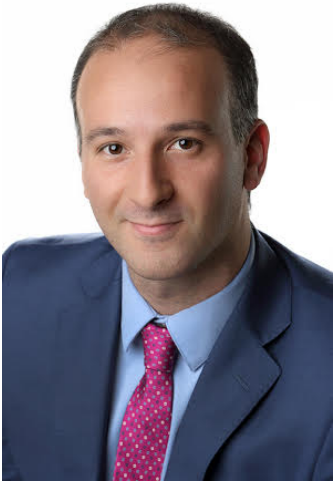

Emilio Ferrara was a Research Scientist at the Indiana University Network Science Institute. Currently he is a Computer Scientist at the University of Southern California's Information Sciences Institute.

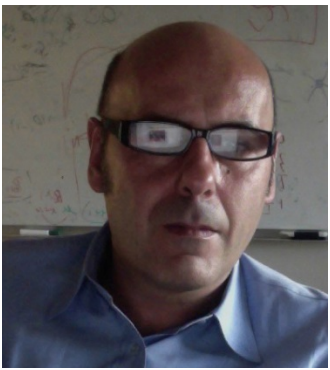

Alessandro Flammini is an associate professor in the School of Informatics and Computing at Indiana University.

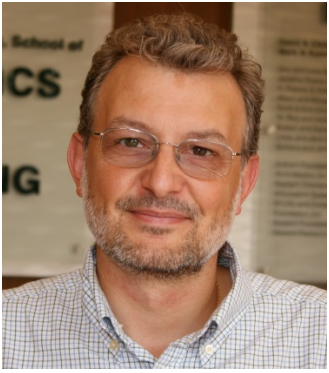

Filippo Menczer is a Professor of Informatics and Computer Science and the Director of the Center for Complex Networks and Systems Research at Indiana University, Bloomington.